\documentclass[letterpaper, 10 pt, conference]{ieeeconf}  

\IEEEoverridecommandlockouts                              

\usepackage{array}
\usepackage{graphicx}
\usepackage{cite}
\usepackage{amsfonts,amssymb}  
\usepackage[cmex10]{amsmath}

\newtheorem{thm}{Theorem}[section]

\newtheorem{prob}[thm]{\bf Problem}
\newtheorem{df}[thm]{\bf Definition}

\title{\LARGE \bf
Optimal Mission Planner with Timed Temporal Logic Constraints 
}

\author{Yuchen Zhou, Dipankar Maity and John S. Baras
\thanks{*This work is supported by US AFOSR MURI grant FA9550-09-1-0538, NSF grant CNS-1035655, NIST grant 70NANB11H148 and by DARPA (through ARO) grant W911NF1410384. }
\thanks{The authors are with the Department of Electrical and Computer Engineering, and the Institute for Systems Research, University of Maryland, College Park, Maryland, USA. email: {\tt\small \{yzh89, dmaity, baras\}@umd.edu}}
}

\begin{document}

\maketitle
\thispagestyle{empty}
\pagestyle{empty}

\begin{abstract}
In this paper, we present an optimization based method for path planning of a mobile robot subject to time bounded temporal constraints, in a dynamic environment. Temporal logic (TL) can address very complex task specification such as safety, coverage, motion sequencing etc. We use metric temporal logic (MTL) to encode the task specifications with timing constraints. We then translate the MTL formulae into mixed integer linear constraints and solve the associated optimization problem using a mixed integer linear program solver. 
This approach is different from the automata based methods which generate a finite abstraction of the environment and dynamics, and use an automata theoretic approach to formally generate a path that satisfies the TL. We have applied our approach on several case studies in complex dynamical environments subjected to timed temporal specifications.
\end{abstract}


\section{Introduction}

Autonomous aircraft have been deployed for agriculture research and management, surveillance and sensor coverage for threat detection and disaster search and rescue operations. All these applications require the tasks to be performed in an optimal manner with specific timing constraints. The high level task specifications for these applications generally consist of temporal ordering of subtasks, motion sequencing and synchronization etc. Given such specifications, it is desirable to synthesize a reference trajectory that is both optimal considering the dynamics of the vehicle and satisfies the temporal constraints. Motion planning \cite{latombe}, \cite{lavalle}, at its early stage, considered optimal planning to reach a goal from an initial position while avoiding obstacles \cite{choset}. New techniques such as artificial potential functions \cite{choset}, \cite{xi}, cell decomposition and probabilistic roadmaps  \cite{lavalle} are introduced for efficient planning in complex environment \cite{Sharma}, \cite{LavSharma} and high dimensional state-space. However, these approaches failed when task specifications have multiple goals or specific ordering of goals, for example surveying some areas in particular sequence.

Temporal logic \cite{baier, Clarke, Quottrupi} provides a compact mathematical formulation for specifying such complex mission specifications. Previous approaches mainly focus on the usage of linear temporal logic (LTL), which can specify tasks such as visiting goals, periodically surveying areas, staying stable and safe. The main drawback of the LTL formulation is that it cannot specify time distance between tasks. In surveillance examples, a simple task may be to individually monitor multiple areas for at least $x$ amount of time. Additionally, the LTL formulation commonly assumes the environment to be static. Traditional approaches commonly start with creating a finite abstraction of the environment including the dynamics, then combine it with the automata that is generated from the LTL specification \cite{Pappas}. The cell decomposition performed in the abstraction process requires the environment to be static; but in most situations this is not the case. For example, the use of Unmanned Aerial Vehicles (UAVs) for surveillance in the commercial airspace needs to consider motion of other aircraft. The other weakness of the automata-based approach is that it is computationally expensive.  In this work, we are interested in motion planning for surveillance in an airspace with finite time task constraints and safety guarantees. For this work, we only consider the other aircraft in the target area as dynamic obstacles to the UAV. Further, we assume that the motions of these dynamic obstacles can be either predicted during the planning or are known a priori.  

Due to the limitations of the previous approaches, we instead present the method based on metric temporal logic (MTL) \cite{MTL}, \cite{MTL1} and an optimization problem formulation to solve the planning problem. MTL extends the LTL \cite{baier} temporal operators so that it can express the requirements on time distance between events and event durations. This allows us to describe the dynamic obstacle and survey durations in our case study. 

An optimization based method for LTL was previously proposed and extended by \cite{KaramanCDC,WolffICRA}. In \cite{KaramanCDC},  the authors propose to transform LTL specifications to mixed integer constraints and solve the planning problem for finite horizon using either a mixed integer linear program (MILP) or a mixed integer quadratic program (MIQP) solver. In \cite{WolffICRA}, the algorithm is extended to infinite horizon so that trajectories can contain loops. However, none of the methods consider dynamic environment or moving obstacles, time varying constraints, or duration of the tasks. MTL was used as a temporal constraint to a routing planning problem in \cite{MTLRounting}. Their formulation unfortunately does not allow users to incorporate dynamics of the vehicle. 

In this paper, we consider a path planning problem for surveillance under survey durations constraints for each region and overall temporal constraint to visit each region within given times. Our problem is considerably different from the problems formulated in the existing literature in the sense that we not only consider dynamic environment but also associate each subtask with a duration constraint. We generate a path that guarantee safety by avoiding static and moving obstacles in the workspace and the path is optimal in the sense that it minimizes a predefined cost function. We do not adopt the linear encoding from \cite{WolffICRA}, since moving obstacles is not periodic in nature. Similar to their approaches, we adopt the usage of mixed logic dynamic (MLD) to model vehicle dynamics, so that the overall problem is a MILP (considering linear cost function).

Our main contribution is usage of MTL to specify time bounded tasks for the mission planner and reformulation of the problem into a MILP. We also demonstrate the methods on the case studies of using a quadrotor and ground vehicle to survey multiple areas given the MTL specifications. The rest of the paper is organized as follows. In section \ref{sec:pre} we present the fundamentals of MTL and define the overall motion planning problem. We then formulate it into a mixed integer linear optimization problem incorporating the temporal constraints in section \ref{sec:prob}. Afterward, we demonstrate our approach on motion planning for different simulation setups. 

\section{Preliminaries}\label{sec:pre}

In this paper we consider a surveying task in an area by a robot whose dynamics are given by the nonlinear model (\ref{eqn1}).
  \begin{equation} \label{eqn1}
   x(t+1)=f(t,x(t),u(t))
  \end{equation}
where  $x(t) \in \mathcal{X}$, $x(0) \in \mathcal{X}_0 \subseteq \mathcal{X}$, and $u(t) \in \mathcal{U}$ for all $t=0,1,2, \cdots$. 
  Let us denote the trajectory of the system (\ref{eqn1}) starting at $t_0$ with initial condition $x_0$ and input $u(t)$ as $\mathbf{x}^{x_0,u}_{t_0}=\{x(s)~| s\geq t_0 ~~ x(t+1)=f(t,x(t),u(t)), ~ x(t_0)=x_0\}$. 
  For brevity, we will use $\mathbf{x}_{t_0}$ instead of $\mathbf{x}^{x_0,u}_{t_0}$ whenever we do not need the explicit information about $u(t)$ and $x_0$.
\begin{df}
   \textit{An atomic proposition is a statement about the system variables (x) that is either $\mathbf{True} (\top)$ or $\mathbf{False} ( \perp)$ for some given values of the state variables.} \cite{KaramanCDC}
  \end{df}
	
 Let $\Pi =\{ \pi_1, \pi_2, \cdots \pi_n \}$ be 
the set of atomic propositions which labels $\mathcal{X}$ as a collection of survey areas, free space, obstacles etc.
The moving obstacles make the free space to change from time to time, and hence labeling 
    the environment is time dependent. We define a map that labels the time varying environment as follows
    \begin{equation} \label{fdef}
     L: \mathcal{X} \times \mathcal{I} \rightarrow 2^{\Pi}
    \end{equation}
    where $\mathcal{I} =\{[a,b]| \hspace{.25 cm} b  > a \geq 0 \} $ and $2^{\Pi}$ is denoted as the power set of $\Pi$. In general, $\mathcal{I}$ 
    is used to denote a time interval but it can be used to denote a time instance as well. 
    
    The trajectory of the system is a sequence of states such that each state $x(t)$ stays in $\mathcal{X}$ for all $t$ and there exists $u(t) \in \mathcal{U}$ for all 
  $t$ such that $x(t+1) =f(t,x(t),u(t))$. 
   Corresponding to each trajectory $\mathbf{x}_{0}$, the sequence of atomic proposition satisfied is given by $\mathcal{L}(\mathbf{x}_{0})=L(x(0),0)L(x(1),1)...$
 
 The high level specification of the surveying task will be expressed formally using MTL which can incorporate timing specifications. 
  
 \subsection{Metric Temporal Logic (MTL)}

Metric temporal logic, a member of temporal logic family, deals with model checking under timing constraints. The formulas for MTL are build on atomic propositions by obeying some grammar.
\begin{df} \label{def1}
 \textit{The syntax of MTL formulas are defined according to the following grammar rules:}
 \begin{center}
 $\phi ::= \top ~| ~\pi~ |~\neg \phi~ | ~\phi \vee \phi ~|~\phi \mathbf{U}_I \phi  ~ $
 \end{center} 
 \end{df}
 where $I\subseteq [0, \infty]$ is an interval with end points in $\mathbb{N} \cup \{\infty\}$. $\pi \in \Pi$,
$\top$ and $\bot(=\neg\top)$ are the Boolean constants $true$ and $false$ respectively. $\vee$ denotes the disjunction operator and $\neg$ denotes the negation operator. 
$\mathbf{U}_I$ symbolizes the timed Until operator. Sometimes we will represent $\mathbf{U}_{[0, \infty]}$ by $\mathbf{U}$. 
Other Boolean and temporal operators such as conjunction ($\wedge$),  eventually within $I$ ($\Diamond_I$), always on $I$ ($\Box_I$) etc. can be represented using the grammar desired in definition \ref{def1}. For example,
we can express time constrained eventually operator $\Diamond_I\phi \equiv \top \mathbf{U}_I\phi$ and so on.

\begin{df}\label{ltlsym}
 \textit{The semantics of any MTL formula $\phi$ is recursively defined over a trajectory $x_t$ as:\\
 $x_{t} \models \pi$ iff $\pi \in L(x(t),t)$\\
 $x_{t} \models \neg \pi$ iff $\pi \notin L(x(t),t)$\\
 $x_{t} \models \phi_1\vee \phi_2$ iff $x_{t} \models \phi_1$ or $x_{t} \models \phi_2$\\
 $x_{t} \models \phi_1\wedge \phi_2$ iff $x_{t} \models \phi_1$ and $x_{t} \models \phi_2$\\
 $x_{t} \models \bigcirc \phi$ iff $x_{t+1} \models \phi$ \\
 $x_{t} \models \phi_1\mathbf{U}_I \phi_2$ iff $\exists s \in I$ s.t. $x_{t+s} \models  \phi_2$ and $\forall$ $s' \leq s, ~ x_{t+s'} \models \phi_1$.}

\end{df}
Thus, the expression $\phi_1 \mathbf{U}_I \phi_2$ means that $\phi_2$ will be true within time interval $I$ and until $\phi_2$ becomes true $\phi_1$ must be true. 
Similarly, the release operator $\phi_1 \mathbf{R} \phi_2$ denotes the specification that $\phi_2$ should always hold and it can be released only when $\phi_1$ is true. 
The MTL operator $\bigcirc \phi$ means that the specification $\phi$ is true at next time instance, $\Box_I \phi$  means that $\phi$ is always true for the time duration $I$, 
$\Diamond_I \phi$ means that $\phi$ will eventually become true within the time interval $I$. Composition of two or more MTL operators can express very sophisticated 
specifications; for example $\Diamond_{I_1} \Box_{I_2} \phi$ means that within time interval $I_1$, $\phi$ will be true and from that instance it will hold true always for a duration
of $I_2$. Other Boolean operators such as implication ($\Rightarrow$) and equivalence ($\Leftrightarrow$) can be expressed using the grammar rules and semantics given in definitions 
\ref{def1} and \ref{ltlsym}. More details on MTL grammar and semantics can be found in \cite{MTL}.  
Satisfaction of a temporal specification $\phi$ by a trajectory $\mathbf{x}_{t_0}$ will be denoted as $\mathbf{x}_{t_0} \models \phi$. 

\section{Problem Formulation and Solution} \label{sec:prob}
 
 We consider a planning problem to periodically survey some selected areas in a given workspace. There are specific time bounds, associated with the regions, by which the surveillance has to be finished. Given the system dynamics (\ref{eqn1}), the objective is to find a suitable control law that will steer the robot in the survey area so that all regions are surveyed within the time bound and that control will optimize some cost function as well. The surveying task and its associated timing constraints and safety constraints can be expressed formally by Metric Temporal Logic (MTL). Let $\phi$ denote the MTL formula for the surveying task, and $J(x(t,u),u)$ be a cost function to make 
the path optimal in some sense. We formally 
present our planning objective as an optimization problem given in Problem (\ref{optproblem})
\begin{prob}\label{optproblem}
\[\begin{array}{c c} \underset{u}{\min} \text{ }   &~J(x(t,u),u(t)) \\ \text{subject to  } &x(t+1)=f(t,x(t),u(t)) \\ &\mathbf{x}_{t_0} \models \phi\\ \end{array}\]
\end{prob}
In the following section we are going to discuss the linearization techniques for the dynamics of the robot, and our approach to translate an MTL constraint as linear constraints. 

\subsection{Linearized Dynamics of the Robot}

Since we are interested in solving the planning problem as an optimization problem with mixed integer and linear constraints, we need to represent the dynamics of the robot as a linear constraint to the optimization problem (\ref{optproblem}). We will consider surveying some given areas by ground robots as well as aerial robots. The dynamics of the autonomous aerial robot and ground robot are given in (\ref{quadrotor}) and (\ref{car}) respectively. 
\subsubsection{Quadrotor Model}
To capture the dynamics of the quadrotor properly, we need two coordinate frames. One of them is a fixed frame 
and will be named as earth frame, and the second one is body frame which moves with the quadrotor. The transformation matrix from body frame to earth frame is $R(t)$. 
The quadrotor dynamics has twelve state variables ($x,y,z,v_x,v_y,v_z,\phi,\theta,\psi,p,q,r$), where $\xi=[x,y,z]^T$ and $v=[v_x,v_y,v_z]^T$ represent the position and velocity of the 
quadrotor w.r.t the body frame. $(\phi,\theta,\psi)$ are the roll, pitch and yaw angle, and $\Omega=[p,q,r]^T$ are the rates of change of roll, pitch and yaw respectively.  

The Newton-Euler formalism for the quadrotor rigid body dynamics in earth fixed frame is given by: 
\begin{align} \label{quadrotor}
 \dot{\xi} & =v  \nonumber \\
 \dot{v} & = -g\mathbf{e_3} + \frac{F}{m}R\mathbf{e_3}\\ \nonumber
 \dot R &=R \hat{\Omega} \\ \nonumber
 \dot\Omega &= J^{-1} (-\Omega \times J\Omega + u)
\end{align}
where $g$ is the acceleration due to gravity, $\mathbf{e_3}=[0, 0, 1]^T$, $F$ is the total lift force and $u=[u_1,u_2,u_3]^T$ are the torques applied. $F$ and $u$ are the control inputs. More details on the quadrotor dynamics can be
found in \cite{Kumar},\cite{Garcia}.
For this work, we linearize the dynamics (\ref{quadrotor}) about the hover with yaw constraint to be zero, as it has been done in \cite{WolffICRA}. Since $\psi$ is constrained to be zero, 
we remove $\psi$ and $r$ from our system and make the system ten dimensional. Consequently, we only need three control inputs, $F, u_1$, and $u_2$ for the system. The linearized model is the same as what is done in  \cite{WolffICRA}, \cite{Dustin}. The system matrices for the linearized
 model are: \\
 $A=\begin{bmatrix} \mathbf{0} & I & \mathbf{0} & \mathbf{0} \\ 
 \mathbf{0} & \mathbf{0} & \begin{bmatrix} 0  & g \\ -g & 0\\0 & 0 \end{bmatrix} & \mathbf{0}\\ 
 \mathbf{0} &\mathbf{0} &\mathbf{0} & I \\ \mathbf{0} &\mathbf{0} &\mathbf{0} &\mathbf{0}\\
 \end{bmatrix}; ~ B = \begin{bmatrix} \mathbf{0} & \mathbf{0}\\ \begin{bmatrix} 0\\  0 \\ 1/m \end{bmatrix} &\mathbf{0} \\ \mathbf{0} & \mathbf{0} \\\mathbf{0} & I_{2 \times 3} J^{-1}\\ \end{bmatrix}$
 $I_{2,3}= \begin{bmatrix}  1 & 0 &0\\0 &1 &0 \\\end{bmatrix}$ \\
 All zero and identity matrices in $A$ and $B$ are of proper dimensions.  
\subsubsection{Car-Like Model}
We also investigate our approach on a car-like dynamical system (\ref{car}). The system has three state variables: positions ($x,y$), heading angle $\theta$. 

\begin{equation}\label{car}
 \left[ \begin{array}{c} \dot{x} \\ \dot{y} \\ \dot{\theta} \end{array} \right] = \left[ \begin{array}{c c} \cos(\theta) & 0 \\ \sin(\theta) & 0 \\ 0 & 1 \end{array} \right] \left[ \begin{array}{c} u_1 \\ u_2 \end{array}\right]
\end{equation}
where $u_1$ and $u_2$ are the control inputs. We linearize this nonlinear model about several values for $\theta$ and depending on the value of $\theta$, the closest linearization was used to 
drive the linearized system. The linearization is similar to what is suggested in \cite{WolffICRA}. The linearized system matrices at $\hat{\theta}$ are given by:

$A=\begin{bmatrix} 0 & 0 & -\hat u_1 \sin(\hat{\theta})\\ 0 &0& \hat u_1 \cos(\hat{\theta})\\ 0 &0 &0 \\\end{bmatrix};  ~B= \begin{bmatrix} \cos(\hat{\theta}) & 0\\ \sin(\hat{\theta}) &0\\ 0 & 1 \end{bmatrix}$

\subsection{Mixed Integer Linear Constraints}

In this section, we demonstrate our approach to translate a time-bounded temporal logic formula to linear constraints on state variables and inputs. The easiest example of it would be how to 
express the temporal constraint that $x(t)$ lies within a convex polygon $\mathcal{P}$ at time $t$. This simple example will serve as a building block for other complicated temporal operators.
Any convex polygon $\mathcal{P}$ can be represented as an intersection of several halfspaces. A halfspace is expressed by a set of points, $\mathcal{H}_i=\{ x: ~ h_i^Tx\leq k_i\}$. 
Thus, $x(t) \in \mathcal{P}$ is equivalent to $x(t) \in \cap_{i=1}^n \mathcal{H}_i =\cap_{i=1}^n \{ x: ~ h_i^Tx \leq k_i \}$. The temporal constraint that $x(t)$ will be inside $\mathcal{P}$
 for all $t \in \{t_1, t_1+1, \cdots t_1+n \}$ can be represented by the set of linear constraints $\{ h_i^Tx(t)\leq k_i \}$ for all $ i =\{1,2,\cdots, n\}$ and $\forall t\in \{t_1, t_1+1, \cdots t_1+n \}$.
 
  We adopted a method similar to the method used in \cite{KaramanCDC} to translate temporal constraint $\phi$ into mixed integer linear constraints. We extend it to incorporate duration for task completion and loop constraints of the trajectory. 
  Comparing to \cite{WolffICRA}, we only enforce the loop constraints at the trajectory level instead of the temporal logic level, since the moving obstacles is not periodic in nature. The planning process will be repeated when the vehicle returns to the initial point.

In a polygonal environment, atomic propositions (AP), $p \in \Pi$, can be related to states of the system using disjunction and conjunction of halfspaces. In other words, the relationship between 
measured outputs such as location of the vehicle and the halfspaces defines the proposition used in the temporal logics. Consider the convex polygon case and let $z_i^t \in \{0,1\}$ be the binary variables associated with halfspaces $\{x(t) : h_i^T x(t)\leq k_i\}$ at time $t=0,...,N$. 
We enforce the following constraint $z_i^t=1$ if and only if $h_i^T x(t) \leq k_i$ by adding the linear constraints,
\begin{equation} 
 h_i^T x(t) \leq k_i + M(1-z_i^t) 
\end{equation} 
\[ h_i^T x(t) \geq k_i -M z_i^t +\epsilon\]
where $M$ is a large positive number and $\epsilon$ is a small positive number. If we denote $P_t^\mathcal{P}=\wedge_{i=1}^n z_i^t$, then $P_t^{\mathcal{P}}=1$ if and only if $x(t) \in \mathcal{P}$. This can be extended to the nonconvex case by decomposing the polygon to convex ones and linking them using disjunction operators. As discussed later in this section, the disjunction operator can also be translated to mixed integer linear constraints.

We will use $F_\phi(x, z, u, t)$ to denote the set of all mixed integer linear constraints corresponding to the temporal
logic formula $\phi$. Once we have formulated $F_p(x,z,u,t)$ for atomic propositions $p$, we can find $F_\phi (x,z,u,t)$ for any MTL formula $\phi$. 
The next essential part of the semantics of MTL is the Boolean operations, such as $\neg$, $\wedge$, $\vee$. Similarly these operators can be translated into linear constraints. 
Let $t \in \{0,1,...,N\}$, $P_t^\phi$ be the continuous variables within [0,1] associated with formula $\phi$ made up with propositions $p \in \Pi$ at time $t$.
\begin{itemize}
	\item $\phi=\neg p$ is the negation of an atomic proposition, and it can be modeled as 
	\begin{align} \label{nega} 
	P_t^{\phi} = 1- P_t^p. 
	\end{align}
	\item The conjunction operation, $\phi = \wedge^m_{i=1}p_i$, is modeled as 
	\begin{align}
	P_t^{\phi} & \leq P_t^{p_i}, \quad i=1,...,m , \\ \nonumber 
	P_t^{\phi} & \geq 1-m+ \sum_{i=1}^m {P_t^{p_i}},  
	\end{align}
	\item  The disjunction operator, $\phi = \vee^m_{i=1}p_i$, is modeled as 
	\begin{align}
	P_t^{\phi} & \geq P_t^{p_i}, \quad i=1,...,m , \\ \nonumber 
	P_t^{\phi} & \leq \sum_{i=1}^m {P_t^{p_i}},  
	\end{align}
\end{itemize}

Similarly, the temporal operators can be modeled using linear constraints as well. Let $t \in \{0,1,...,N-t_2\}$, where $[t_1,t_2]$ is the time interval used in the MTL. 
\begin{itemize}
	\item Eventually: $\phi=\Diamond_{[t_1,t_2]} p$ is equivalent to
	\begin{align}
	P_t^{\phi} & \geq P_\tau^{p}, \quad \tau \in \{t+t_1,...,t+t_2 \} \\ \nonumber
	P_t^{\phi} & \leq \sum_{\tau=t+t_1}^{t+t_2} {P_\tau^{p}} 
	\end{align}
	\item Always: $\phi=\Box_{[t_1,t_2]} p$ is equivalent to
	\begin{align}
	P_t^{\phi} & \leq P_\tau^{p}, \quad \tau \in\{t+t_1,...,t+t_2\}\\ \nonumber
	P_t^{\phi} & \geq \sum_{\tau=t+t_1}^{t+t_2} {P_\tau^{p}} -(t_2-t_1),
	\end{align}
	\item Until: $\phi=p \mathcal{U}_{[t_1,t_2]} q$ is equivalent to
	\begin{align}\label{eq:until}
	a_{tj} & \leq P_q^{j} \quad j \in \{t+t_1,\cdots,t+ t_2\}, \nonumber \\ \nonumber
	a_{tj} & \leq P_p^{k} \quad k \in \{t, \cdots, j-1\}, j \in \{t+t_1,\cdots,t+ t_2\}, \\ \nonumber
	a_{tj} & \geq P_q^{j} + \sum_{k=t}^{j-1} P_p^k -(j-t) \quad j \in \{t+t_1,\cdots,t+ t_2\},\\
	P_t^{\phi} & \leq \sum_{j=t+t_1}^{t+t_2}a_{tj}, \\ \nonumber
	P_t^{\phi} & \geq a_{tj} \quad j \in \{t+t_1,\cdots, t+t_2\}. \nonumber
	\end{align}
	%
\end{itemize}
The until formulation (\ref{eq:until}) is obtained similarly to \cite{KaramanCDC}. For MTL, it is modified by noticing the following equality,
\[P_t^{\phi}= \bigvee_{j=t+t_1}^{t+t_2}\left((\wedge_{k=t}^{k=j-1}P_p^k)  \wedge P_q^j \right). \]  
Other combinations for different temporal operators are also straight forward and we are not enumerating them for the sake of space, but one can easily derive them by using (\ref{nega}) through (\ref{eq:until}). 

Using this approach, we translate the given high level specification in MTL ($\mathbf{x}_{t_0} \models \phi$) to a set of mixed integer linear constraints $F_\phi (x,z,u,t)$. At the end, we add the constraint $P_0^\phi=1$, i.e. the overall specification $\phi$ is satisfied. Since Boolean variables are only introduced when halfspaces are defined, the computation cost of MILP is at most exponential to the number of halfspaces times the discrete steps $N$.
%
%

\section{Case Study and Discussion}
We apply our method for solving mission planning with finite time constraints on two different workspaces. Both workspaces contain static and moving obstacles.The experiments are run through YALMIP-CPLEX on a computer with 3.4GHz processor and 8GB memory. We performed the simulation for both a quadrotor model and a car model. 
\begin{figure}[t]
\centering
\includegraphics[width=2.5in]{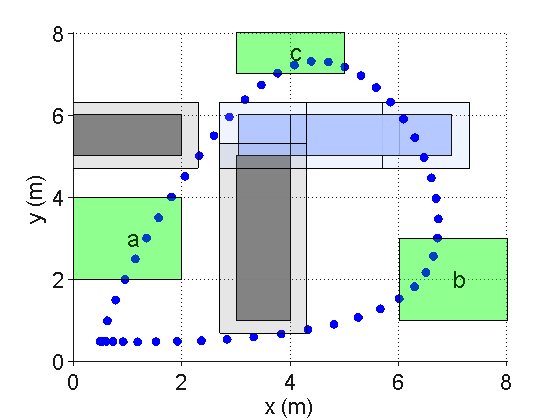}
\caption{Workspace setup of the first test case. Blue area represents an obstacle moving from right to left. Grey areas represent static obstacles. Green areas represent survey areas used in the temporal logic. The shaded areas around obstacles represent boundary of the obstacles for discrete planning, so that the obstacles can be avoided even for continuous dynamics. The resulting 2D trajectory of the quadrotor for $\phi_1$ is shown as blue dots. The resulting motion is counter clockwise. }
\label{fig:proj_1}
\end{figure}
\begin{figure}[t]
\centering
\includegraphics[width=2.5in]{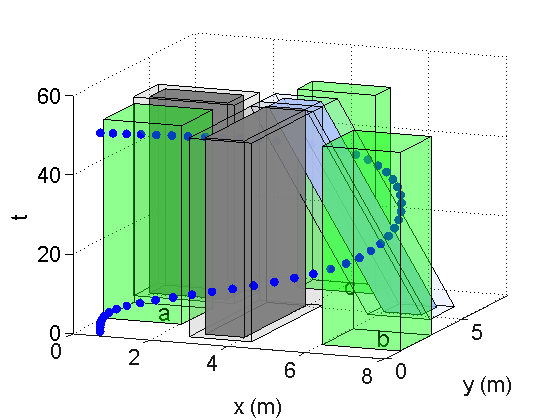}
\caption{Time-state-space representation of the environment and resulting trajectory for the quadrotor with temporal constraints $\phi_1$. The vehicle starts from (0.5,0.5) and surveys area b, c and a sequentially.}
\label{fig:timespace_1}
\end{figure}

The first environment is the one shown in Fig. \ref{fig:proj_1}, where blue and gray areas represent moving and static obstacles respectively, and green areas represent survey targets. 

Let the temporal specification be the following 
\[ \phi_1= \Diamond \Box_{[0,2]} A \wedge \Diamond \Box_{[0,2]} B \wedge \Diamond \Box_{[0,2]} C \wedge \Box \neg O 
\]
\[\phi_2= \Diamond \Box_{[0,2]} A \wedge \Diamond \Box_{[0,2]} B \wedge \Diamond \Box_{[0,2]} C \wedge \Box \neg O \wedge \neg B\, \mathcal{U}_{[0,N]} A
\]
where $O$ represents the static and moving obstacles, $N$ is the horizon of the planning trajectory. Such $N$ can be generally obtained by performing a feasibility test using MILP solver starting from $N=2$, and increasing $N$ until finding a feasible one. For the purpose of comparing different temporal constraints, we choose a feasible horizon $N=50$ for both MTL specifications $\phi_1$ and $\phi_2$. The specification $\phi_1$ requires the vehicle to visit areas $A$, $B$ and $C$ eventually and stay there for at least 2 time units, while avoiding obstacles $O$. $\phi_2$ adds an additional requirement on the ordering between $A$ and $B$, so that region $A$ has to be visited first. We consider the cost function, $J$, to be minimized is $\sum_{t=0}^{N}|u(t)|$. The dynamics of the vehicles are discretized at a rate of 2Hz.

\begin{figure}[t]
\centering
\includegraphics[width=2.25in]{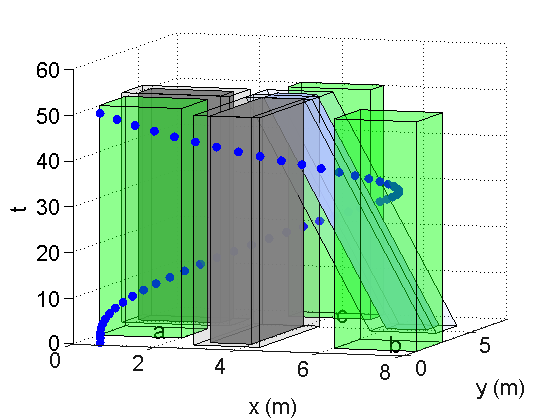}
\caption{Time-state-space representation of the environment and resulting trajectory for the quadrotor with temporal constraints $\phi_2$. Because of the additional ordering requirement, the vehicle covers area a first before visiting area b.}
\label{fig:timespace_3}
\end{figure}

The resulting trajectory for the quadrotor with temporal specification $\phi_1$ is plotted in time-state-space in Fig. \ref{fig:timespace_1}. The projection of the trajectory on the workspace is shown in Fig. \ref{fig:proj_1}. The motion of the quadrotor in Fig. \ref{fig:proj_1} is counter clockwise. 
The quadrotor safely avoids the moving obstacle by navigating through the area after the obstacle passes. The survey duration in individual area also satisfies the requirements as shown in Fig. \ref{fig:proj_1}. The quadrotor stays within each survey area for 5 discrete time units which is equivalent to 2s duration. These plots show a better realization of visiting individual areas comparing to \cite{KaramanCDC} and \cite{WolffICRA}. In their results, trajectory visits targeted areas for only 1 discrete time unit which is not desired for most surveillance tasks. Additional local planning is possible to generate a sweeping pattern so that the target areas can be covered by onboard sensors in designed durations.

The resulting trajectory for the vehicle with specification $\phi_2$ is shown in time-state-space in Fig. \ref{fig:timespace_3}. As can be verified from the trajectory, the vehicle travels first to area $A$ before visiting $B$ as specified. The CPLEX solver returns the solution for the first trajectory in 20.8 sec while the second one takes 34.7 sec. The additional continuous variables used in the until encoding have large influence in the performance. One of the possible future research directions would be finding different encoding of the until operator to improve the speed of the algorithm.

\begin{figure}[t]
\centering
\includegraphics[width=2.25in]{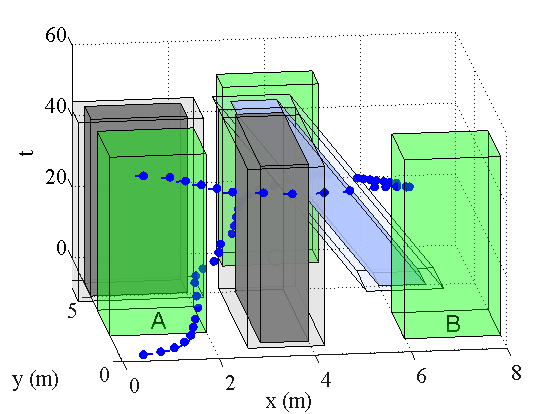}
\caption{Time-state-space representation of the environment and resulting trajectory for the car model with temporal constraints $\phi_1$. The result for $\phi_2$ is similar since the optimal solution for $\phi_1$ already goes through area A first.}
\label{fig:car_1}
\end{figure}

The result of specification $\phi_1$ for the car model is shown in Fig. \ref{fig:car_1}, where the small blue arrow associated with each blue dot indicates the instantaneous heading of the vehicle. The computation time is 150s. The longer computation time is caused by the additional binary variables introduced by the linearization of the dynamics at various heading angles.

\begin{figure}[t]
\centering
\includegraphics[width=2.5in]{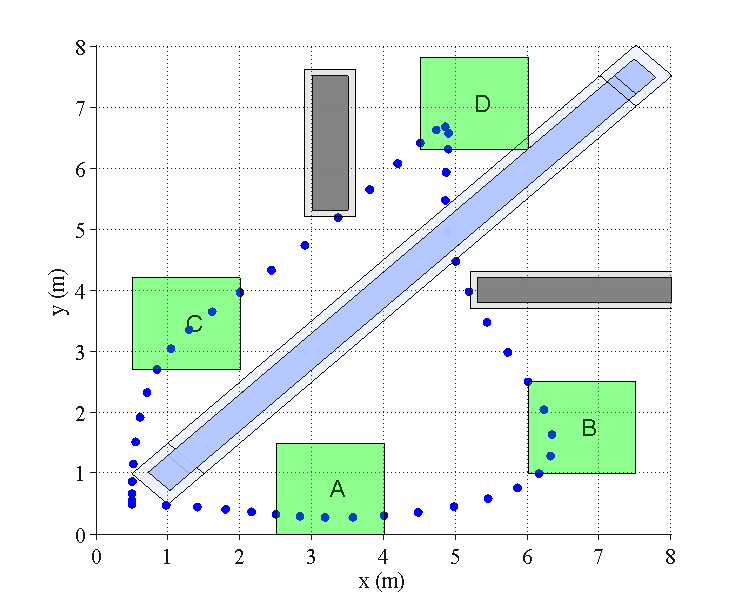}
\caption{Workspace setup of the second test case. The moving obstacle moves from the the bottom left corner to upper right corner. 2D trajectory of the vehicle is shown as blue dots. The resulting motion is clockwise. }
\label{fig:proj_2}
\end{figure}
The second environment considers a fast moving obstacle that moves across the workspace diagonally, and hence the vehicle has to adjust its motion accordingly. The environment is shown in Fig \ref{fig:proj_2}. Similar to the previous example, it shows the motion of the moving obstacle, static obstacles and survey areas. 

The temporal logic specifications are similar to the previous one but have an additional area to be visited. We also tested the case when certain area has to be visited first. The result is similar to previous cases, so we only show the plots for $\phi_3$.
\[ \phi_3= \Diamond \Box_{[0,2]} A \wedge \Diamond \Box_{[0,2]} B \wedge \Diamond \Box_{[0,2]} C \wedge \Diamond \Box_{[0,2]} D \wedge \Box \neg O 
\]

\begin{figure}[t]
\centering
\includegraphics[width=2.25in]{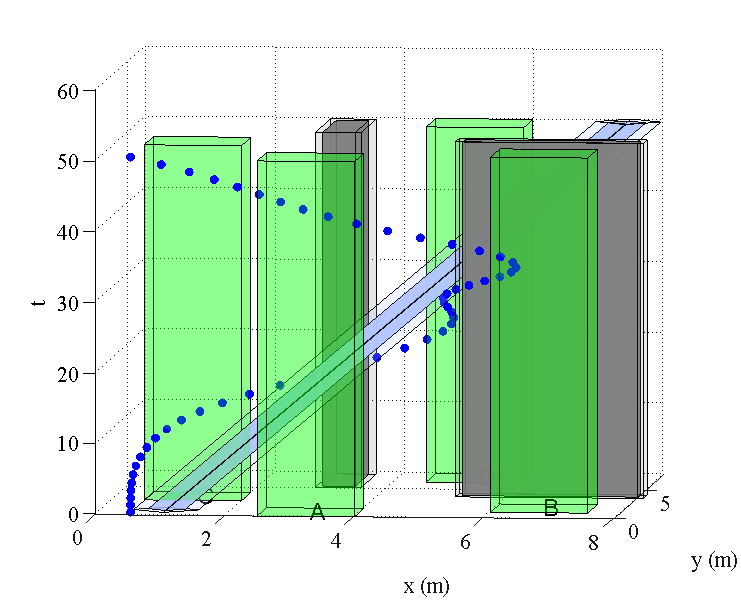}
\caption{Time-state-space representation of the environment and resulting trajectory for the quadrotor with temporal constraints $\phi_3$. The planned trajectory is very close to the dynamic obstacles. }
\label{fig:timespace_6}
\end{figure}

The resulting trajectory in time-state-space for the quadrotor with temporal specification $\phi_3$ is plotted in Fig. \ref{fig:timespace_6}. The projected trajectory on the workspace of the robot is shown in Fig. \ref{fig:proj_2}. The motion of the vehicle is clockwise in Fig. \ref{fig:proj_2}. As can be seen from Fig. \ref{fig:timespace_6}, the quadrotor safely avoids the moving obstacle and the static ones nearby. The survey duration in individual area also satisfies the requirements as shown in Fig. \ref{fig:proj_2}. The computation time is 138.8s. The increase in computation time is because the complex environment introduces more binary variables.
%

\begin{figure}[t]
\centering
\includegraphics[width=2.5in]{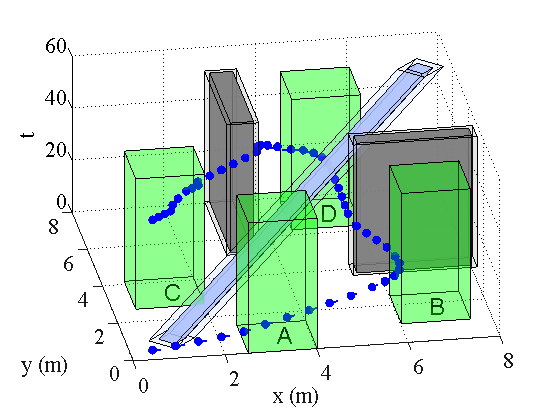}
\caption{Time-state-space representation of the environment and resulting trajectory for the car model with temporal constraints $\phi_3$. The result for $\phi_4$ is the same since the optimal solution already goes through A first.}
\label{fig:car_2}
\end{figure}

The result of specification $\phi_3$ for the car model is shown in Fig. \ref{fig:car_2}, where the blue arrows indicate the heading of the vehicle. The computation time is 500s. 

\section{Conclusion}

In this paper, we have presented an optimization based approach to plan the trajectory of a robot in a dynamic environment to perform some temporal task with finite time constraints. Our approach is simple in the sense that it translates the time constraints and the temporal specifications as linear constraints on the state and input variables. We have linearized the dynamics of the robot in order to formulate the problem as a Linear Programming problem. 
We have considered polygonal environments for our case studies, but if the environment is not polygonal, one can approximate it with a polygonal environment. We have used a binary variable ($z$) with each halfspace, so if the polygonal approximation of the environment contains too many halfspaces, the problem would be complex. 

Due to space constraints, we have reported the case studies only for quadrotor and car dynamics. The simulation results show promising performance of our approach to find an optimal solution. We consider dynamic but deterministic environments and uncertainties in the dynamics of the robot are also not considered. There are many possible directions such as planning in uncertain environment, stochastic dynamics of the robots or multi-robot cooperative planning that one might consider as an extension of this framework.







\bibliographystyle{IEEEtran}
\bibliography{bib}
%



\end{document}